\documentclass[aps,twocolumn, prd, amsmath, floats,floatfix, superscriptaddress, nofootinbib, showkeys]{revtex4}
\usepackage{amssymb}
\usepackage{amsmath}
\usepackage{verbatim}
\usepackage{mathrsfs}
\usepackage{amsfonts}
\usepackage{latexsym}
\usepackage{epsfig}
\usepackage{color}
\usepackage{graphicx,subfigure}
\usepackage{units}
\usepackage{hyperref}
\usepackage{envmath}
\usepackage{natbib}
\usepackage{ctable}
\makeatletter \newcommand*{\rom}[1]{\expandafter\@slowromancap\romannumeral #1@} \makeatother
\makeatother
\DeclareUnicodeCharacter{2212}{-}
\DeclareUnicodeCharacter{060C}{!!!FIX ME!!!}
\begin{document}
\definecolor{orange}{rgb}{0.9,0.45,0}
\def\CovDev{D}
\def\Res{{\mathcal R}}
\def\Gammaflat{\hat \Gamma}
\def\metricflat{\hat \gamma}
\def\Dflat{\hat {\mathcal D}}
\def\part_n{\partial_\perp}
%
\def\Lie{\mathcal{L}}
\def\A{\mathcal{X}}
\def\Aphi{\A_{\phi}}
\def\hAphi{\hat{\A}_{\phi}}
\def\E{\mathcal{E}}
\def\Ham{\mathcal{H}}
\def\M{\mathcal{M}}
\def\R{\mathcal{R}}
\def\p{\partial}
\def\hg{\hat{\gamma}}
\def\hA{\hat{A}}
\def\hD{\hat{D}}
\def\hE{\hat{E}}
\def\hR{\hat{R}}
\def\hcA{\hat{\mathcal{A}}}
\def\hDelt{\hat{\triangle}}
\def\na{\nabla}
\def\dif{{\rm{d}}}
\def\non{\nonumber}
\newcommand{\erf}{\textrm{erf}}
%
\renewcommand{\t}{\times}
\long\def\symbolfootnote[#1]#2{\begingroup%
\def\thefootnote{\fnsymbol{footnote}}\footnote[#1]{#2}\endgroup}
\title{New Accretion Constraint on the Evaporation of Primordial
Black Holes}

\author{Seyed Sajad Tabasi} 
\email{sstabasi98@gmail.com}
\affiliation{Department of Physics, Sharif University of Technology, P. O. Box 11155-9161, Tehran, Iran}
\affiliation{PDAT Laboratory, Department of Physics, K. N. Toosi University of Technology, P.O. Box 15875-4416, Tehran, Iran}
\author{Mahsa Berahman}
\email{mahsa.berahman@email.kntu.ac.ir}
\affiliation{Department of Physics, K.N. Toosi University of Technology, P.O. Box 15875-4416, Tehran, Iran}
\affiliation{PDAT Laboratory, Department of Physics, K. N. Toosi University of Technology, P.O. Box 15875-4416, Tehran, Iran}
\author{Javad T. Firouzjaee}
\email{firouzjaee@kntu.ac.ir}
\affiliation{Department of Physics, K.N. Toosi University of Technology, P.O. Box 15875-4416, Tehran, Iran}
\affiliation{PDAT Laboratory, Department of Physics, K. N. Toosi University of Technology, P.O. Box 15875-4416, Tehran, Iran}
\affiliation{School of Physics, Institute for Research in Fundamental Sciences (IPM), P.O. Box 19395-5531, Tehran, Iran}

\begin{abstract}
In this paper, we have investigated the processes of evaporation and accretion of primordial black holes during the radiation-dominated era and the matter-dominated era. This subject is very important since usually these two processes are considered independent of each other. In other words, previous works consider them in such a way that they do not have a direct effect on each other, and as a result, their effects on the mass of primordial black holes are calculated separately. The calculations of this paper indicate that assuming these two processes independently of each other will lead to wrong results that only give correct answers within certain limits. In fact, in general, it is a mistake to consider the static state for the event horizon of primordial black holes and perform calculations related to their evaporation, while the radius of primordial black holes is constantly changing due to accretion. In addition, we have shown that considering the dynamic event horizon in some masses and in some times can lead to the shutdown of the Hawking evaporation process. This study is much more accurate and detailed than our previous study. These calculations show well the mass evolution of primordial black holes from the time of formation to the end of the matter-dominated era, taking into account both the main processes governing black holes, evaporation and accretion.

 
\end{abstract}

\maketitle

\section{Introduction}
 The detection of gravitational waves generated by the mergers of two black holes  \cite{LIGOScientific:2016aoc, LIGOScientific:2021djp} has led to renewed interest in Primordial Black Holes (PBHs) \cite{Sasaki:2018dmp,Carr:2020gox,Green:2020jor},  as they could be part of a fraction of the events observed by the LIGO/Virgo/KAGRA collaboration \cite{Hutsi:2020sol,DeLuca:2021wjr, Franciolini:2021tla}.

PBHs may be formed through the gravitational collapse of rare overdense regions upon horizon entry in the early stages of the universe's evolution. The collapse could take place during the radiation-dominated era when PBHs are generated only if the initial amplitude of the density perturbation is on the far side of a large threshold (see, e.g., \cite{Niemeyer:1999ak,Shibata:1999zs,Allahyari:2016osl, Musco:2020jjb}). 

There are two main features of PBHs dynamics first, their evaporation by Hawking radiation, and second their accretion, which is due to the nature of the black hole's significant gravity. PBHs Hawking radiation flux is not independent of its accretion flux \cite{Firouzjaee:2014zfa,Firouzjaee:2015bqa,Firouzjaee:2015wps}.
Since all stationary BHs evaporate due to Hawking radiation \cite{Hawking:1975vcx}, losing their mass in a time related to their initial mass by equation $\tau \sim M^3$, then the PBH with initial mass less than $ 10^{15} g $ have entirely evaporated until now. 
With respect that the accretion could overcome Hawking radiation during the radiation-dominated era and causes PBHs radius to grow, therefore the constrain from evaporation for PBHs is reduced from $ 10^{15} g $ to $ 10^{14} g $ \cite{Tabasi:2021cxo}. Therefore, they safely show the constraints down to $ M\geq 10^{14} g $, which leads to the remained possible PBHs mass range windows to be extended foe explaining dark matter. 

These two dynamical features help us to know the abundance of PBHs which share in detected gravitational waves and dark matter mass fraction.
The abundance of PBHs is constrained by observations in different mass ranges (for a comprehensive review, see \cite{Carr:2020gox}). 

For example, Ricotti, Ostriker, and Mack \cite{Ricotti:2007au} derived strong constraints from the cosmic microwave background (CMB) frequency spectrum and temperature and polarization anisotropies for PBHs more massive than one solar mass. The basic idea about these constraints is that PBHs accrete primordial gas in the early universe and then convert a fraction of the accreted mass to radiation which affects the CMB. To proceed, first one has to model the PBHs accretion to quantify their mass value in time. Second, the type of the accretion flux (gas) and the era of the universe in which the PBHs evolve in it determine PBHs mass spectrum.

The rest of this paper is organized as follows. In section \rom{2} we have an overview of some general cosmological equations. Then, in section \rom{3}, we have explained the evaporation process and the equations leading it, and then we have continued the same process for the accretion process. In section \rom{4}, the equations of accretion of matter and radiation are analyzed. In section \rom{5}, considering the significance of the two eras, radiation-dominated and matter-dominated, we examined the evolution of mass due to the accretion of matter and radiation in each era separately. The mass graph is drawn in terms of time, and the effects of accretion of radiation and matter are discussed. In section \rom{6}, we have talked about the presence or absence of evaporation by examining the rate of increase in the radius of the PBH due to accretion.

\section{General Equations}
All PBHs have been formed in the radiation-dominated era. The study of PBHs mass gives much information about its evolutionary process and effects on the surrounding environment. The mass of PBHs that formed at the time t after the Big Bang is equivalent to or less than the Hubble mass \cite{Carr:2009jm}

\begin{equation}
\label{e1}
M_{PBH} \sim \frac{c^3t}{G}\sim 10^{15}(\frac{t}{10^{-23}s})g,
\end{equation}

where $c\simeq3\times10^{8}m/s$ is the speed of light and $G\simeq6.67\times10^{-11}m^{3}/kgs^2$ is the gravitational constant.

The cosmological evolution of PBHs, such as accretion, evaporation, and merging, can significantly impact PBHs mass and release radiation, injecting energy into the surrounding medium, strongly affecting its thermal state, and leaving influential observable signatures \cite{Villanueva-Domingo:2021spv}.

To study of PBHs, we need to survey the universe's evolution. Friedmann equations describe the homogeneous and isotropic universe as \cite{Rice:2017avg}

\begin{equation}
\label{e2}
(\frac{\dot{a}}{a})^2+\frac{kc^2}{a^2}=\frac{8\pi G}{3}\rho,
\end{equation}

\begin{equation}
\label{e3}
\frac{\ddot{a}}{a}=-\frac{4\pi G}{3}(\rho+\frac{3P}{c^2}),
\end{equation}

and the total energy conservation equation is

\begin{equation}
\label{e4}
\dot{\rho}+3H(\rho+p)=0.
\end{equation}

General equation of state is $p = \omega\rho$, where $\omega$ for matter, radiation and cosmological constant are 0, 1/3, −1, thus we can rewrite the Eq.\eqref{e4} as \cite{Nayak:2011sk}

\begin{equation}
\label{e5}
\begin{split}
\rho=\rho_{cr}(\frac{a}{a_{0}})^{-3(1+\omega)}\hspace{1cm}\\
\vspace{0.5cm}
\rho(a)\propto 
\begin{cases}
        a^{-4} & \hspace{0.5cm} \text{Radiation}\\
        a^{-3} & \hspace{0.5cm} \text{Matter} \\
        constant & \hspace{0.5cm}   \text{Vacuum}
    \hspace{1 cm}
\end{cases}
\end{split}
,
\end{equation}

and by substituting Eq.\eqref{e3} in Eq.\eqref{e5} we have

\begin{equation}
\label{e6}
\hspace{1.2cm}a(t)\propto \begin{cases}
        t^{\frac{1}{2}} & \hspace{0.5cm} \text{Radiation} \\
        t^{\frac{2}{3}} & \hspace{0.5cm} \text{Matter} \\
        e^{H_{0}t} & \hspace{0.5cm} \text{Vacuum}
    \end{cases}.
\hspace{2 cm}
\end{equation}

Now we want to calculate the rate of mass change of PBHs through evaporation and accretion processes.
\subsection{Evaporation}
After inspecting quantum properties for black holes, Hawking indicated that black holes emit particles with a thermal spectrum \cite{hawking-75}. The properties of the emitted particles depend on mass, angular momentum, and charge of BHs \cite{Cheek:2021odj}. 
We consider PBHs as Schwarzschild black holes \cite{Carr:2009jm}

\begin{equation}
\label{e7}
\frac{dM_{PBH}}{dt}=-f_{eva}4\pi R_{PBH}^{2}c\rho_{r},
\end{equation}

where $f_{eva}$ is the evaporation efficiency factor, $R_{PBH}$ is the PBH radius, and $\rho_{r}$ is radiation density. Evaporation efficiency factor plays a vital role in the evaporation rate, and its value depends upon PBHs physical parameters and environment. The function $\rho_{r}$ is given by \cite{Rice:2017avg}

\begin{equation}
\label{e8}
\rho_{r}=\frac{\pi^2}{30}g_{*}(T_{PBH})\frac{(T_{PBH}k_{B})^4}{{\hbar}^3c^5},
\end{equation}
where $g_{*}$ is the number of relativistic particle degrees-of-freedom which is obtained by 
\begin{equation}
\label{e9}
g_{*}(T_{PBH})=\sum_{i}(\omega_{i}g_{i}).
\end{equation}

In order to get a numerical value for $g_{*}(T_{PBH})$, we need to have values of $\omega_{i}$ and $g_{i}$

\begin{equation}
\label{e10}
\begin{split}
\omega_{i}=
    \begin{cases}
        2s_{i}+1 & \hspace{0.5cm} massive\hspace{2mm} particles\\
        2 & \hspace{0.5cm} massless \hspace{2mm} species \\
        1 & \hspace{0.5cm}   s_{i}=0
    \end{cases}\\
\vspace{0.3cm}
\hspace{1.3cm} g_{i}(T_{PBH})=
    \begin{cases}
        1.82 & \hspace{0.5cm}   s=0 \\
        1.0 & \hspace{0.5cm}   s=\frac{1}{2}\\
        0.41 & \hspace{0.5cm}   s=1\\
        0.05 & \hspace{0.5cm}  s=2\\
    \end{cases}\hspace{1.5cm}
\end{split},
\end{equation}

and obviously, $s_{i}$ is the particle spin. 
Hence, if $M_{PBH}\ll10^{11} g$ for standard model particles $g_{*}(T_{PBH})\simeq 108$. As a substitute, the Minimal Supersymmetric Standard Model (MSSM) approximates $g_{*}(T_{PBH})\simeq 316$ \cite{Hooper:2020otu,Keith:2020jww}.

Although we have used Eq.\eqref{e7}  to continue, it can be rewritten by Eq.\eqref{e8} as follows

\begin{equation}
\label{e11}
\frac{dM_{PBH}}{dt}=-\frac{8\pi^3}{15}\frac{f_{eva}g_{*}}{4}\frac{M_{PBH}^2}{c^5\hbar c M_{Pl}^4}(k_{B}T_{PBH})^4.
\end{equation}

In this equation, $T_{PBH}$ is the temperature of the radiation particles from PBHs, which is equal to PBHs temperature. As we will demonstrate, PBHs temperature is critical in the accretion and evaporation processes that are given as \cite{Carr:2009jm}

\begin{equation}
\label{e12}
T_{PBH}=\frac{\hbar c^3}{8\pi G k_{B}M_{PBH}} \simeq10^{-7}(\frac{M_{PBH}}{M_{\odot}})^{-1}.
\end{equation}

This process slowly reduces the PBH mass, so if the dominant process is evaporation, the lifetime of a PBH with initial mass M derives from the following equation \cite{Chisholm:2011kn}

\begin{equation}
\label{e13}
\tau(M)\simeq (10^{-26} s)(\frac{M}{1 g})^3.
\end{equation}

\subsection{Accretion}
As mentioned, accretion has a significant effect on the evolution of PBHs. Infalling matter and photons onto PBHs increase the mass and other observable parameters. 

The physical parameter of a cosmological fluid determines the accretion rate in each cosmic epoch \cite{Rice:2017avg}. In this study, we focus on accretion equations, and all calculations are performed by considering spherical symmetric condition. The Bondi-Hoyle accretion model is used for this goal \cite{Bondi:1952ni}

\begin{equation}
\label{e14}
\frac{dM_{PBH}}{dt}=4\pi R_{PBH}^2\rho v.
\end{equation}

In accretion of radiation $v={c}/{\sqrt{3}}$ and $R_{PBH}=R_{s}={2GM_{PBH}}/{c^2}$. Therefore we can rewrite Eq.\eqref{e14} as follows

\begin{equation}
\label{e15}
\frac{dM_{PBH}}{dt}=16\pi G^2 M_{PBH}^2\rho_{r} (\frac{c}{\sqrt{3}})^{-3} f_{acc},
\end{equation}

where $f_{acc}$ is the accretion efficiency. Now we consider conditions under which a PBH acquires matter in the accretion process. This case is more complex, and we need more information about the environment. To obtain the rate of mass increases by baryonic matter, we use the following equation

\begin{equation}
\label{e16}
\frac{dM_{b}}{dt}=\lambda4\pi m_{H} n_{gas} v_{eff} r_{B}^2.
\end{equation}

Here, $n_{gas}$ is the number density, $r_{B}=G M_{PBH} v_{eff}^{-2}$ is the Bondi-Hoyle radius and $v_{eff}=(v_{rel}^2+c_{s}^2)^{\frac{1}{2}}$ is the effective velocity of PBH, expressed in terms of the PBH relative velocity $v_{rel}$ with regard to the gas with sound speed $c_{s}$ \cite{Ricotti:2007au}. The gas viscosity, Compton drag, Compton cooling by CMB photons, and free electron fraction are factors that determine the value of dimensionless accretion rate $\lambda$, which is effective in obtaining the final mass value. Provided both Compton drag and Compton cooling are negligible, the classic Bondi problem can be solved for an adiabatic gas \cite{Ali-Haimoud:2016mbv}.

\section{ACCRETION OF UNIVERSE'S COMPONENTS}
As we know, the universe is made up of baryonic matter (gas), dark matter, radiation, and dark energy. In this section, we will examine the accretion of these components. However, due to the fact that in this paper, we study equations until the end of the matter-dominated era, and we expect that in these two eras, mass gain by matter and radiation will be dominant, our focus will be on the accretion of matter and radiation so we neglect accretion of dark energy. In the following section, we peruse these two regimes of accretion individually.

\subsection{Accretion of radiation}
The presence of CMB anisotropies and fluctuations on scales larger than the Hubble radius in the recombination era point strongly to the early inflationary epoch \cite{Brandenberger:2012zb}. The thermal bath result from reheating is an essential aspect of inflation. Thereupon we can consider the universe is the precise black-body \cite{Allahverdi:2005fq}. Considering that equations associated with accretion of radiation are distinct in the radiation-dominated era and matter-dominated era, we discuss them separately. In the radiation-dominated era, photons from thermal bath fall into PBHs and increase their mass. As mentioned, we consider spherical symmetrical accretion and use Eq.\eqref{e15}. We need to have this equation in terms of time or redshift to examine the evolution of PBHs. Therefore, by using Eq.\eqref{e5}, we know $\rho_{r}=\rho_{cr}(a/ a_{0})^{-4}$ then Eq.\eqref{e6} is used to enter the time parameter, and Eq.\eqref{e15} is rewritten as follows \cite{Nayak:2011sk}

\begin{equation}
\label{e17}
\begin{split}
\frac{dM_{PBH}}{dt}=16\pi G^2 \rho_{cr} \Omega_{r}^0 (\frac{c}{\sqrt{3}})^{-3} f_{acc}\\
\times(t_{1}^{-\frac{2}{3}} t_{2}^\frac{8}{3} e^{-4H_{0}(t_{2}-t_{0})}) (\frac{M_{PBH}}{t})^2
\end{split},
\end{equation}

where $\rho_{cr}=9.2\times10^{-30}{g}/{cm^3}$ is the critical energy density, $\Omega_{r}^0=4.2\times10^{-5}$ is relative contribution of relativistic particles, $t_{1}=2.1\times10^{12}s$ is the time of end of the radiation-dominated era, $t_{2}=2.4\times10^{17}s$ is the time of end of the matter-dominated era, and $t_{0}=4.4\times10^{17}s$ is the present time \cite{Rubakov:2017xzr}. By solving the differential equation of Eq.\eqref{e17}, the final mass of PBHs due to accretion of radiation in radiation-dominated is obtained in terms of time

\begin{equation}
\label{e18}
\begin{split}
M_{ R-RD}(t)=(\frac{1}{M_{i}}+1.3\times10^{-35}f_{acc}(\frac{1}{t}-\frac{1}{t_{i}}))^{-1}\\
for \hspace{0.5cm}t_{i}<t<t_{1}\hspace{2cm}
\end{split},
\end{equation}

where $M_{i}$ is the initial mass, and $t_{i}$ is the formation time of the PBH. Explicitly, Eq.\eqref{e18} determines the mass resulting from the accretion of radiation at any time during the radiation-dominated era and specifically the final mass of the PBH at the end of this era. Correspondingly in the matter-dominated era $\rho_{r}=\rho_{cr}({a}/{a_{0}})^{-3}$ and $a(t)\propto t^{{2}/{3}}$ so we have

\begin{equation}
\label{e19}
\begin{split}
\frac{dM_{PBH}}{dt}=16\pi G^2 \rho_{cr} \Omega_{r}^0 (\frac{c}{\sqrt{3}})^{-3} f_{acc}\\ \times(t_{2}^{-\frac{8}{3}} e^{-4H_{0}(t_{2}-t_{0})}) \frac{M_{PBH}^2}{t^{\frac{8}{3}}}
\end{split}.
\end{equation}

Now, by solving the differential equation of Eq.\eqref{e19}, we can have mass evolution through the accretion of radiation in the matter-dominated era. The final mass because of accretion of radiation in matter-dominated era is obtained as

\begin{equation}
\label{e20}
\begin{split}
M_{R-MD}(t)=(\frac{1}{M_{i}}+3.5\times10^{-27}f_{acc}(\frac{1}{t^{\frac{5}{3}}}-\frac{1}{t_{1}^{\frac{5}{3}}}))^{-1}\\for \hspace{0.5cm} t_{1}<t<t_{2}\hspace{3cm}
\end{split}.
\end{equation}

In Ref.\cite{Tabasi:2021cxo}, the importance and consequences of correctly determining the value for the accretion efficiency factor have been well studied. However, in the literature, values between 0.05 and 0.2 are usually attributed to it. In all the calculations of this paper, we have considered the value of 0.1 for it.

\begin{figure*}[t]
\label{fig1}
\begin{center}
\subfigure[]{
  \includegraphics[width=65mm]{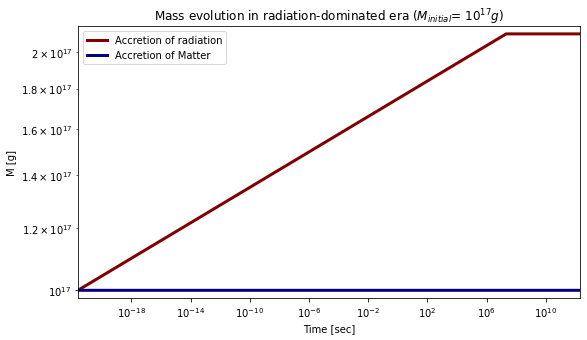}
}
\subfigure[]{
  \includegraphics[width=65mm]{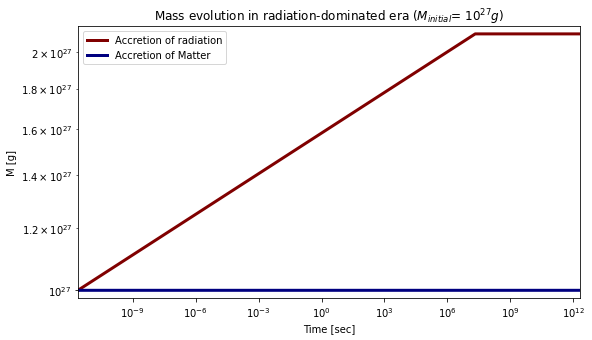}
}
\hspace{0mm}
\subfigure[]{
  \includegraphics[width=65mm]{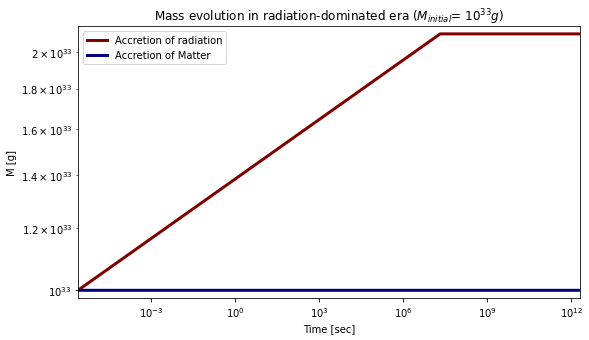}
}
\subfigure[]{
  \includegraphics[width=65mm]{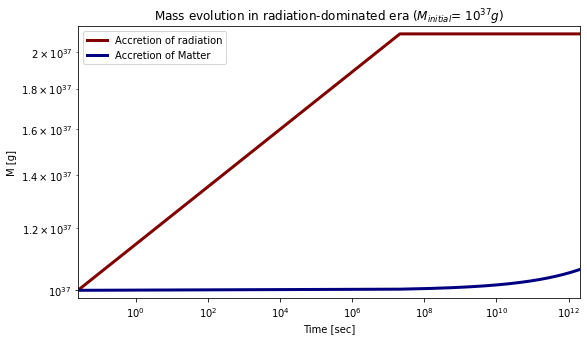}
}
\end{center}
  \vspace*{-3mm}
\begin{center}
\caption{\begin{footnotesize} This figure shows how the mass grows during the radiation-dominated era by accretion of radiation (red line) and accretion of matter (blue line). As expected, during this epoch, mass growth occurs mainly due to accreting of radiation. The graphs are plotted for different initial masses (a)$10^{17}g$,
 (b)$10^{27} g$, (c)$10^{33} g$, and (d)$10^{37} g$. These masses were chosen because there are no strict constraints on these masses to explain for at least part of the dark matter
 .\end{footnotesize}}
 \end{center}
\end{figure*}
\begin{figure*}[b!]
 \label{fig:2}
\hspace{3mm}
\subfigure[]{
  \includegraphics[width=60mm,height=38mm]{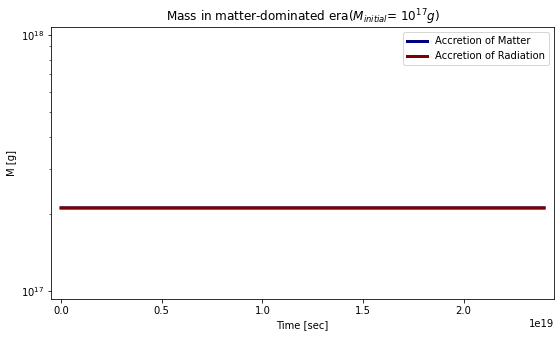}
}
\subfigure[]{
  \includegraphics[width=65mm,height=38mm]{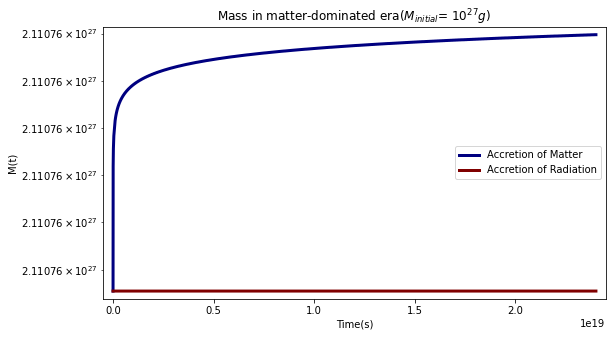}
}
\hspace{0mm}
\subfigure[]{
  \includegraphics[width=65mm,height=38mm]{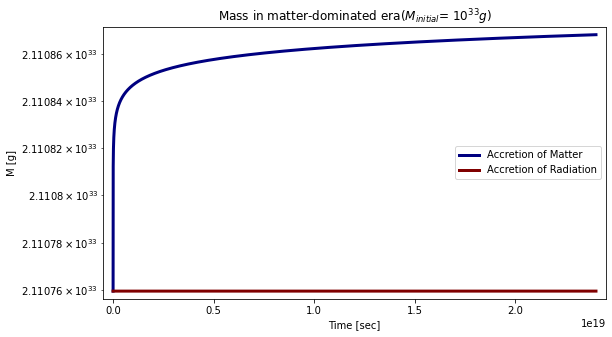}
}
\hspace{2.4mm}
\subfigure[]{
  \includegraphics[width=63mm,height=38mm]{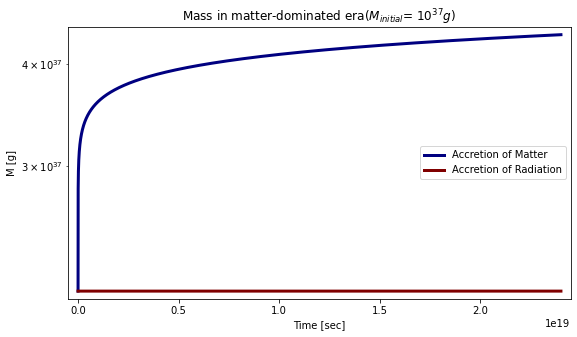}
}
  \vspace*{-3mm}
 \begin{center}
\caption{\begin{footnotesize} In this figure we can see that in the matter-dominated era, the accretion of matter is dominant over the accretion of radiation. In (a) it is obvious that changing the mass is so slight and neither accreting of radiation nor accreting of matter is not significant. Nevertheless, in (b),(c), and (d) accreting of matter evidently change the mass of PBHs.\end{footnotesize}}
 \end{center}
\end{figure*}

\subsection{Accretion of Matter}
Throughout this paper, we assume that the PBH with point mass M is immersed in the Hydrogen gas. In order to continue, we need to refer to Eq.\eqref{e16} and investigate each term of the equation. The numerical value of the mean cosmic gas density is

\begin{equation}
\label{e21}
n_{gas}\simeq 200 cm^{-3} (\frac{1+z}{1000})^3.
\end{equation}

Aforementioned, $v_{eff}$ is a variety of $c_{s}$ and the $v_{rel}$ between the PBH and the medium averaged with a Gaussian distribution. From \cite{Yang:2021agk}
 we have

\begin{equation}
\label{e22}
\sqrt{\langle{v_{L}^2}\rangle}\simeq min[1,\frac{1+z}{1000}]\times 30 km/s.
\end{equation}

Given the equation $c_{s}=(5.7 km.s^{-1})(T_{gas}/2730)^{1/2}$ , to compute the speed of sound, we need the gas temperature. Before decoupling, we can consider the gas temperature was roughly equal to the CMB temperature. After that, $T_{gas}$ started to decrease adiabatically due to the Hubble parameter. Therefore, the value of $c_{s}$ can be written approximately as follows \cite{Ali-Haimoud:2016mbv}

\begin{equation}
\label{e23}
c_{s}\simeq
\begin{cases}
        (5.7 km.s^{-1})(\frac{1+z}{1000})^{\frac{1}{2}} & \hspace{0.5cm} z\gg132 \\
        1800 km.s^{-1} & \hspace{0.5cm}   z\ll132 \\
    \end{cases}.\hspace{1.2cm}
\end{equation}

Finally, we should introduce

\begin{equation}
\label{e24}
v_{eff}\simeq
\begin{cases}
        c_{s}\mathcal{M}^{\frac{1}{2}}[3\sqrt{\frac{2}{2\pi}B(\frac{3}{2},\frac{3}{2})}]^{-\frac{1}{6}} & \hspace{0.5cm} \mathcal{M}\gg1 \\
         c_{s} & \hspace{0.5cm}   \mathcal{M}\ll1 \\
    \end{cases},\hspace{0.3cm}
\end{equation}

where B(x,y) is the beta function, and $\mathcal{M}$  is defined as $\mathcal{M}\equiv{{\sqrt{\langle{v_{L}^2}\rangle}}/{c_{s}}}$ \cite{Mena:2019nhm}.

The value of $\lambda$ must be determined in terms of redshift. We assume the constant free electron fraction $x_{e}$ is equal to the free electron fraction of background  $\overline{x_{e}}=1$ also, we need the characteristic dimensionless Compton drag rate $\beta$ and Compton cooling rate $\gamma$ as a function of redshift. We can get \cite{DeLuca:2020fpg}

\begin{equation}
\label{e25}
\begin{split}
\beta=(\frac{M}{10^4M_{\odot}})(\frac{z+1}{1000})^{\frac{3}{2}}(\frac{v_{eff}}{5700})^{-3}\\ \times[0.275+1.45(\frac{x_{e}}{0.01})(\frac{1+z}{1000})^{\frac{5}{2}}]
\end{split},
\end{equation}

\begin{equation}
\label{e26}
\gamma=\frac{2 m_{p}}{m_{e}(1+x_{e})}\beta.
\end{equation}

Although $\lambda$ can vary according to how $\gamma$ and $\beta$ relate to each other, in general, the following relationship applies to all redshifts \cite{Ali-Haimoud:2016mbv}

\begin{equation}
\label{e27}
\lambda(\beta,\gamma)\approx\frac{\lambda(\gamma; \beta \ll 1)\lambda(\gamma\gg 1;\beta)}{\lambda_{iso}}.
\end{equation}

In this equation, $\lambda_{iso}=1.12$ in the isothermal case and $\lambda_{ad}=0.12$ in tha adiabatic case. Additionally, $\lambda(\gamma;\beta\ll1)$ is the accretion rate numerical solution for $\beta\ll1$ and arbitrary $\gamma$. Similarly $\lambda(\gamma\gg 1;\beta)$ is the numerical solution for $\gamma\gg1$ and arbitrary $\beta$. Eq.\eqref{e28} and Eq.\eqref{e29} show equations of these special $\lambda$ 

\begin{equation}
\label{e28}
\lambda(\gamma;\beta\ll1)\approx\lambda_{ad}+(\lambda_{iso}-\lambda_{ad})(\frac{\gamma^2}{88+\gamma^2})^{0.22},
\end{equation}

\begin{equation}
\label{e29}
\lambda(\beta;\gamma \gg 1)\approx exp[\frac{4.5}{3+\beta^{\frac{3}{4}}}]\times\frac{1}{(\sqrt{1+\beta}+1)^2}.
\end{equation}

Now we have all the parameters of Eq.\eqref{e16} in terms of redshift, and we can substitute them for getting the mass rate equation. As in the previous section, with placing $1+z=(a_{0}/a)=e^{H_{0}(t_{0}-t_{2})}(t_{2}/t_{1})^{2/3}(t_{1}/t)^{1/2}$ according to Eq.\eqref{e6}, we can solve Eq\eqref{e16} in terms of time and obtain the mass evolution of PBHs in the radiation-dominated era

\begin{equation}
    \label{e30}
\begin{split}
    M_{M-RD}(t)=M_{i}(1+\frac{123}{25\times10^{39}}M_{i}(\sqrt[4]{t_{i}} - \sqrt[4]{t}))^{-1}\\ for\hspace{0.5cm} t_{i}<t<t_{1} \hspace{2.5cm}
    \end{split}.
\end{equation}

Besides, the mass evolution equation in terms of time in the matter-dominated era by using $1+z=(a_{0}/a)=e^{H_{0}(t_{0}-t_{2})}(t_{2}/t)^{2/3}$ is

\begin{equation}
    \label{e31}
    \begin{split}
    M_{M-MD}(t)=M_{t-RD}(1+1.5\times10^{-36}M_{t-RD}\ln{\frac{t_{1}}{t}})^{-1}\\for \hspace{0.5cm} t_{1}<t<t_{2} \hspace{3cm}
    \end{split},
\end{equation}

where $M_{t-RD}$ is the final mass of PBHs because of accretion of radiation and matter at the end of the radiation-dominated era.

\begin{figure*}[b!]
\label{fig:3}
\begin{center}
\subfigure[]{
  \includegraphics[width=65mm]{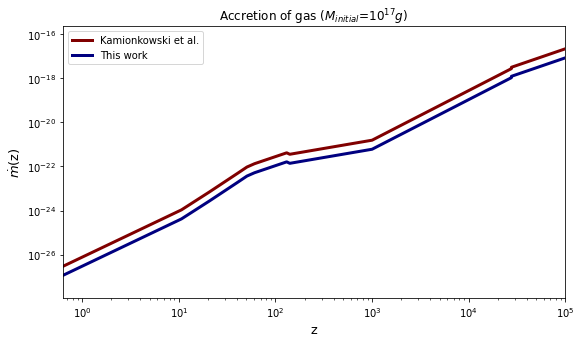}
  }
\subfigure[]{
  \includegraphics[width=65mm]{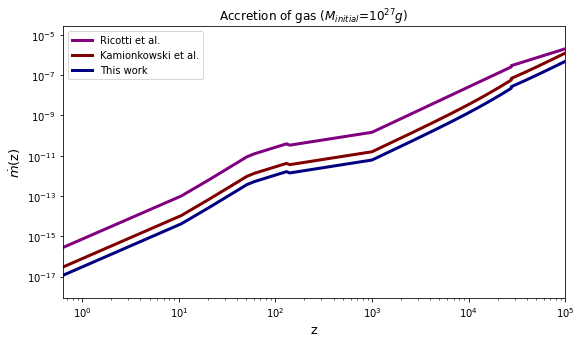}
}
\subfigure[]{
  \includegraphics[width=65mm]{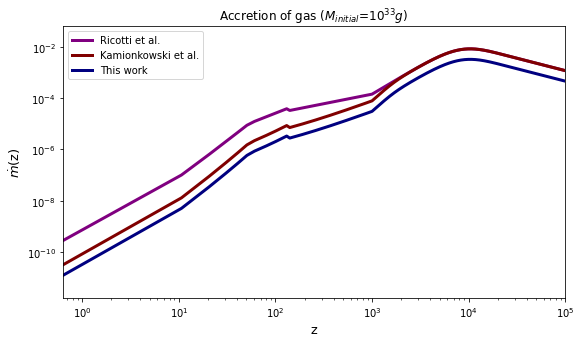}
}
\subfigure[]{
  \includegraphics[width=65mm]{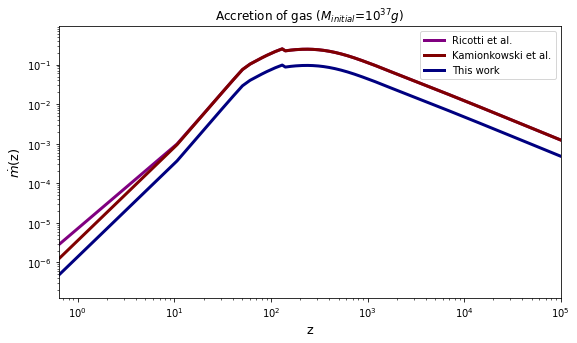}
}
	\end{center}
 \vspace*{-3mm}
\begin{center}
\caption{\begin{footnotesize} We have compared the accretion of gas for selected PBHs masses in our work with results of the papers of Ricotti {\it et al.} and Kamionkowski {\it et al.}. It should be noted that the Ricotti equation does not behave correctly in low masses, such as mass $10^{17 g}$. However, it is clear that all three models are close to each other, and we see similar behavior. Afterward, we can state that our model works correctly.\end{footnotesize}}
\end{center}
\end{figure*}

\begin{figure*}[ht]
 \label{fig:4}
\subfigure[]{
  \includegraphics[width=65mm]{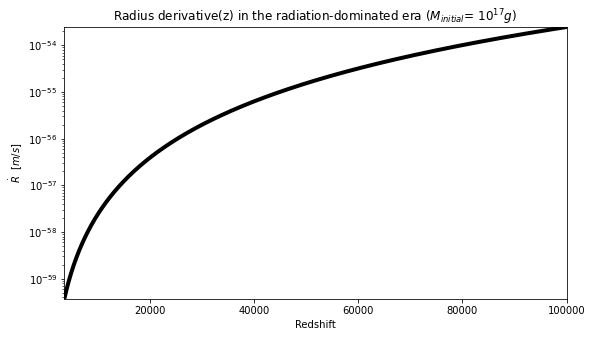}
}
\hspace{0mm}
\subfigure[]{
  \includegraphics[width=65mm]{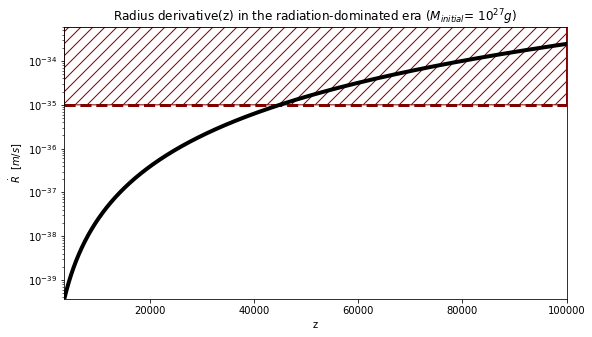}
}
\hspace{0mm}
\subfigure[]{
  \includegraphics[width=65mm]{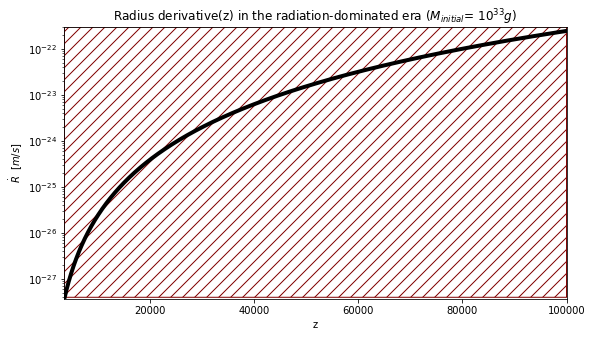}
}
\subfigure[]{
  \includegraphics[width=65mm]{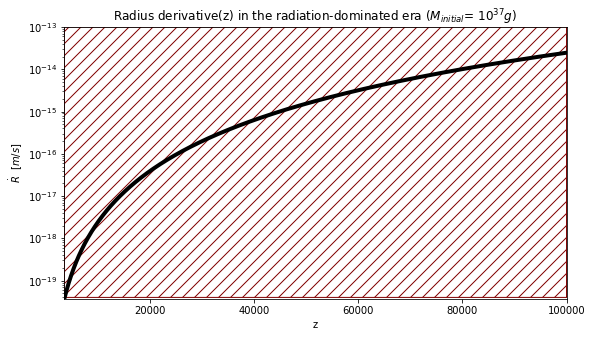}
}
 \vspace*{-3mm}
\begin{center}
\caption{\begin{footnotesize} We have compared the evaporation and the accretion processes in the radiation-dominated era.  In (a), PBH with the initial mass of $10^{17}g$ has low accretion effect, as expected, the evaporation process remains powerful during this period.  (b) shows that in the case of PBHs with the initial mass of $10^{27}g$, evaporation stops early in the radiation-dominated era. (c) and (d) state that the evaporation does not occur in PBHs with high initial mass due to the high growth rate of radius and event horizon.\end{footnotesize}}
\end{center}
\end{figure*}

\begin{figure*}[t]
 \label{fig:5}
\subfigure[]{
  \includegraphics[width=65mm]{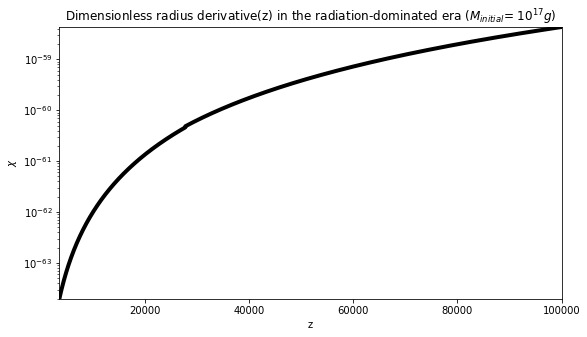}
}
\subfigure[]{
  \includegraphics[width=65mm]{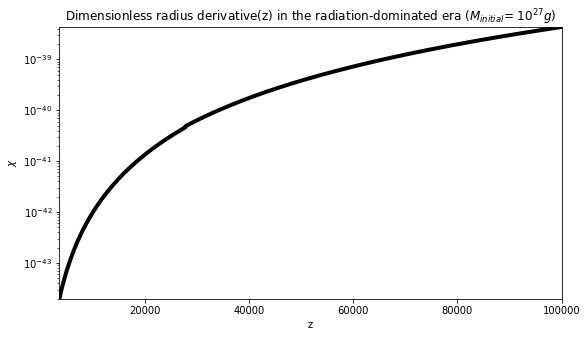}
}
\hspace{0mm}
\subfigure[]{
  \includegraphics[width=65mm]{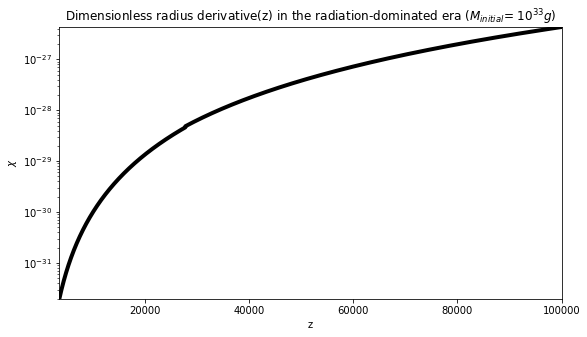}
}
\subfigure[]{
  \includegraphics[width=65mm]{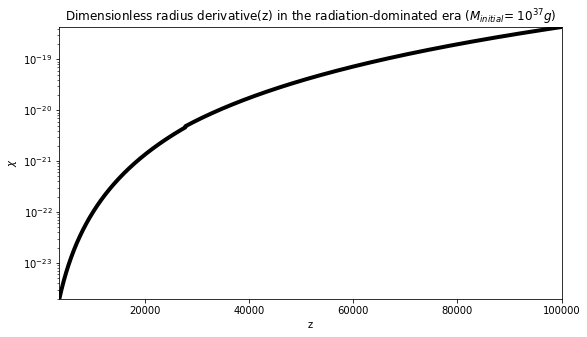}
}
  \vspace*{-3mm}
\begin{center}
\caption{\begin{footnotesize} The changes of dimensionless parameter $\chi$ in the radiation-dominated era for four initial masses (a)$10^{17}g$, (b)$10^{27} g$, (c)$10^{33} g$, and (d)$10^{37} g$ are plotted. To compare this work with future works or other models of accretion, it is very important to pay attention to these plots because calculations related to the type of the accretion, the properties of the cosmic environment and how PBHs were formed are considered in this parameter.\end{footnotesize}}
\end{center}
\end{figure*}

\section{Accretion during the radiation-dominated era}
Radiation and matter fall into PBHs all the time and increase its mass. In the last section, we discussed the evolution of mass for accretion of radiation and accretion of matter. In this section, we want to establish whether the assumptions that radiation has a more serious effect on increasing the mass of the PBH in the radiation-dominated era or the matter is responsible for increasing the mass in the matter-dominated era are correct and not. Due to the importance of observing PBHs, many studies have been conducted on limiting the possible masses for the existence of PBHs and for explaining dark matter. After applying all constrains, including evaporation \cite{Page:1976wx}, lensing \cite{Niikura:2019kqi}, gravitational waves \cite{Raidal:2017mfl}, cosmic microwave background distortions \cite{Kohri:2014lza}, four mass windows $10^{16}–10^{17} g$, $10^{20}–10^{24} g$, and $1–10^4M_{\odot}$ remain \cite{Carr:1997cn,Carr:2021bzv}. In this paper, we have studied a mass from each mass window to analyze the results obtained ($10^{17}$, $10^{27}$, $10^{33}$, and $10^{37} g$), and we have examined the graph of mass over time in two eras separately. Fig.~(1) demonstrates the growth of PBHs mass during radiation-dominated era and compares the effect of matter and radiation on the increase of PBHs mass.

As we expected, radiation during the radiation-dominated era significantly increases the mass of PBHs, and we can neglect the accretion of matter in this epoch.

\section{Accretion during the matter-dominated era}
Now we should investigate the accretion of PBHs during the matter-dominated era. Eq.\eqref{e20} and Eq.\eqref{e30} illustrate the mass of PBHs that has started to devour matter and radiation in this era. Fig.~(2) has satisfied our expectations in the matter-dominated era; the growth of PBHs mass is mainly done because of matter.

To examine the accuracy of our work, we compared our results with the previous works, in particular with  papers of Kamionkowski {\it et al.} and Ricotti {\it et al.} For this goal, it is necessary to define the dimensionless Bondi-Hoyle accretion rate that shows the evolution of the accretion rate normalized to the Eddington rate as $\dot{m}\equiv{{\dot{M_{b}}}/{\dot{M_{Ed}}}}$, where $\dot{M_{Ed}}=1.44\times10^{17}(M_{PBH}/M_{\odot}) \hspace{0.2cm} erg.s^{-1}$ is the Eddington accretion rate. Fig.~(3) gives us practical information about the mass evolution only by gas accretion. This paper uses an analytical solution to calculate the equations as much as possible.

 Regarding our semi-analytical approach, Fig.~(3) depicts a slight difference between the mentioned approach and fully numerical methods. Although in low redshifts, Kamionkwski {\it et al.} have considered the adiabatic accretion in this era because of the neglectable Compton cooling effect, Ricotti {\it et al.} implicitly have assumed that $\gamma \gg 1$ at all times when accounting for Compton drag in the analysis \cite{Ali-Haimoud:2016mbv}.

\begin{figure*}[ht!]
\label{fig:6}
\subfigure[]{
  \includegraphics[width=65mm]{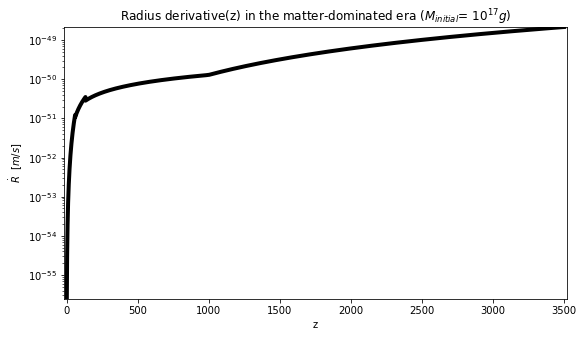}
}
\subfigure[]{
  \includegraphics[width=65mm]{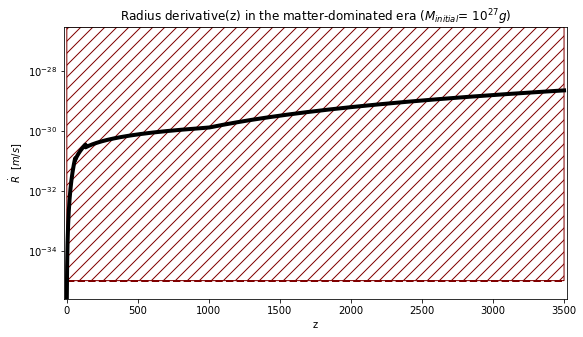}
}
\hspace{0mm}
\subfigure[]{
  \includegraphics[width=65mm]{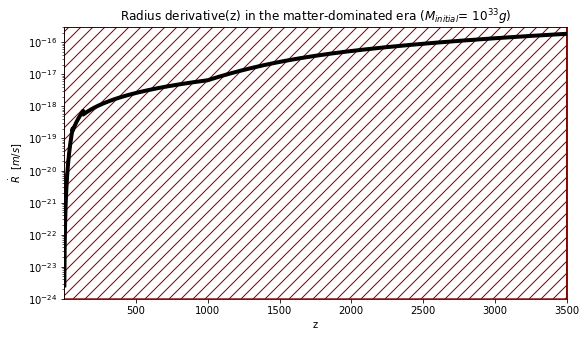}
}
\subfigure[]{
  \includegraphics[width=65mm]{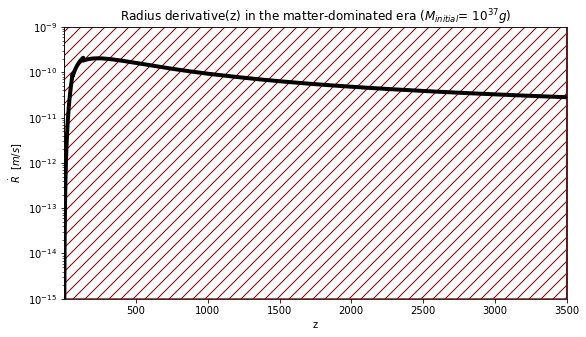}
}
  \vspace*{-3mm}
\begin{center}
\caption{\begin{footnotesize} This figure points out the competition between evaporation and accretion in the matter-dominated era. (a) shows the accretion of PBHs with initial mass $10^{17} g$ that it does not overcome evaporation until the end of the matter-dominated era. PBHs with an initial mass of $10^{27}g$ in (b), after the evaporation turns off, it resumes the evaporation process at the end of the matter-dominated epoch when the change of the radius rate decreases sharply. (c),(d) indicate that radius of PBHs with these masses grows fast that still no evaporation until the end of the matter-dominated era can be seen.\end{footnotesize}}
\end{center}
\end{figure*}

 \begin{figure*}[ht!]
 \label{fig:7}
\subfigure[]{
  \includegraphics[width=65mm]{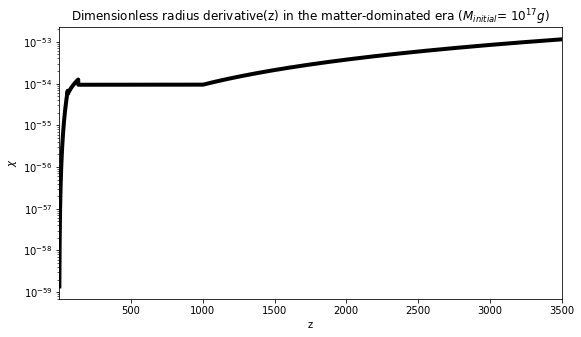}
}
\hspace{0mm}
\subfigure[]{
  \includegraphics[width=65mm]{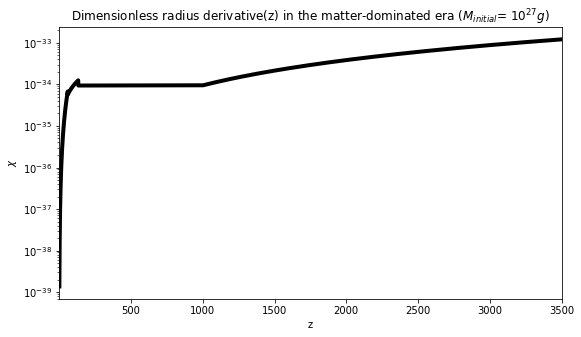}
}
\hspace{0mm}
\subfigure[]{
  \includegraphics[width=65mm]{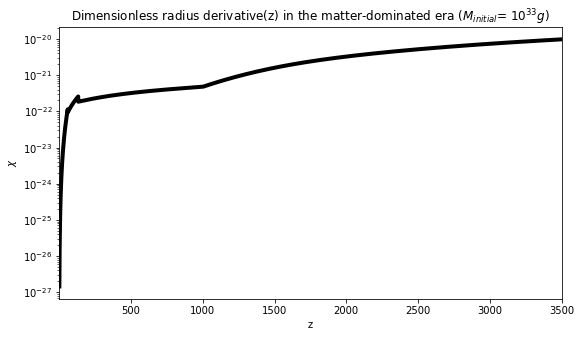}
}
\subfigure[]{
  \includegraphics[width=65mm]{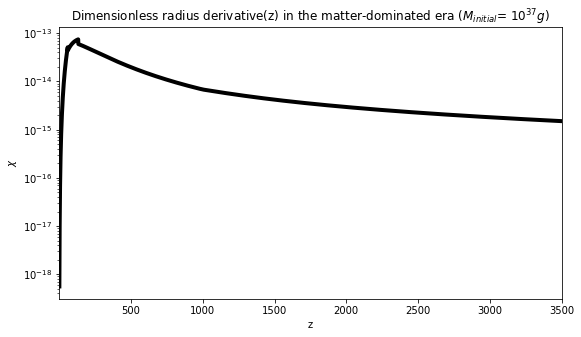}
}
  \vspace*{-3mm}
\begin{center}
\caption{\begin{footnotesize} Similar to the radiation-dominated era the changes of dimensionless parameter $\chi$ in the matter-dominated era for four initial masses (a)$10^{17}g$, (b)$10^{27} g$, (c)$10^{33} g$, and (d)$10^{37} g$ are plotted. Clearly, the behavior of PBHs with the mass $10^{34}g$ is completely different. \end{footnotesize}}
\end{center}
\end{figure*}

\section{Evaporation vs. Accretion}
In previous sections, we investigated the process of PBHs mass increase during the radiation-dominated and matter-dominated era. One of the most meaningful results obtained from a PBH mass is the calculation of its radius. According to Schwarzschild radius relationship, $R_{s}=2GM_{PBH}/c^2$, if we substitute $\dot{M_{b}}$ obtained from the previous parts, we can study Hawking evaporation by comparing the growth rate of the event horizon and Planck length \cite{Tabasi:2021cxo}. 

By using Beckenstein-Hawking entropy $S=A/4l_{p}^2$ and setting characteristics of thermal fluctuations about equilibrium $\delta S \sim 1$, we can estimate the scale of quantum fluctuations of the horizon. We know the horizon is treated as $N\equiv{A/lp^2}$, and equation $\delta A \sim \sqrt{N}\delta a \sim lp \delta r$ holds for each era and radius $r$. Therefore, according to equation $\delta a \sim lp^{2}$ we have $\delta r \sim lp$ \cite{Jacobson:1993hn}. Thus, particles that escape the black hole start their journey from about a Planck length farther than the event horizon. 

The apparent horizon of any dynamical space-time must be inside the event horizon; thus, any virtual pair particle created by vacuum, cannot escape outside and should fall back into PBHs. Due to accretion, a PBH is in the dynamical phase, so it cannot have adiabatic conditions around the apparent horizon for Hawking radiation \cite{Firouzjaee:2015bqa}.

In this context, we are interested in following changing rate of radius for the mentioned masses. We would like to know if for these different masses, there are time periods where evaporation is turned off. In Fig.~(4) and Fig.~(6), radius growth rates are plotted in terms of time for selected masses. In order to facilitate conclusions, the radiation-dominated era and matter-dominated era are separately shown, and the regions where the radius changes are more than the Planck length are crosshatched.
 
As we can see in Fig.~(4), PBHs with the initial mass $10^{17} g$ constantly evaporate during the radiation-dominated era. The situation is a bit more complicated for PBHs with the initial mass of $10^{27}g$. Because according to Fig.~(4).b these PBHs firstly evaporate during the radiation-dominated era, but the radius increase rate shortly exceeds the Planck length and the evaporation process stops. In the case of the other two selected masses, these PBHs do not evaporate at all during radiation-dominated era.

Since it is usually more appropriate to work with dimensionless parameters for comparison, in this paper, we define a new parameter $\chi={\dot{R}}/{v_{eff}}$. In addition to the fact that this parameter is dimensionless and this makes it suitable for comparing different models, there is another reason for defining it. This parameter is dependent on $v_{eff}$ and as a result it is related to sound speed and relative velocity of PBHs. This dependence makes the effects of the cosmic environment which is diverse in various models as well as the relative velocity of the initial PBHs for which there are different estimations to be seen in the changes of this parameter. On the other hand, the type of accretion that is chosen, whether it is spherical symmetrical accretion or dist accretion, also has a serious effect on this parameter. Therefore, the definition of this parameter is necessary. $\chi$ as a function of z for four masses plotted in Fig.~(5) and Fig.~(7).

We apply the same calculations on PBHs during the matter-dominated era. In Fig.~(6), we can see that masses which evaporation process was turned off during the radiation-dominated era do not evaporate in the matter-dominated as well due to the increase in the radius change rate. At the end of the matter-dominated era, the radius change rate drops sharply, whereby quenched evaporation may be reactivated in some masses; for example, in Fig.~(6), we see this condition in the PBH with an initial mass of $10^{27}g$. For the initial mass of $10^{17} g$ g\, radius changes are less than the Planck length and continue to evaporate during this era.

In our opinion, all models in which the evaporation of PBHs has been proposed to justify a phenomenon in the history of the universe should be re-examined. Since the starting and stopping times of Hawking evaporation are different for PBHs with different masses, these calculations must be done first to ensure that PBHs with the proposed masses will evaporate at all at that time or not. This issue is much more important for primordial black holes with low masses. Thus, the calculations related to this work must be checked for them first. We also suggest that the $\chi$ parameter should be used seriously in all future works, because this very important parameter contains many features of a model related to PBHs like the model of their formation, their accretion model, cosmic environment situations, etc. 
\\\\\
\section{Conclusions}
Since PBHs are one of the most important candidates for dark matter, their evolution in time is also very important. As we know, the two main processes that can change the mass of black holes are Hawking radiation and accretion. Therefore, the behavior of these processes must be well understood in order to be able to calculate the evolution of PBHs. The accretion equations can be well represented by the  Bondi-Hoyle model. Of course, this model is a well-defined model with the condition of spherical symmetry. A disk model can also be considered, which will provide more accurate answers. However, for simplicity, the  Bondi-Hoyle model is used in this paper.

On the other hand, Hawking's approach to considering black holes as black bodies and trying to investigate the thermodynamic properties of black holes is very attractive and practical. Although no such radiation has been observed so far, the logic of its existence is so convincing that we cannot deny its existence.
Nevertheless, the main question is whether a black hole can always swallow particles through accretion and emit particles from itself through Hawking evaporation. This question becomes even more important when we realize that any of these processes, when applied to PBHs, can have important cosmological and astrophysical consequences. Thus, without a doubt, this question must be answered.

In this paper, we first showed that in the radiation-dominated era, the rate of mass increase of PBHs due to the swallowing radiation is much higher than the rate of increase of mass due to the swallowing matter. Despite it is the opposite in the matter-dominated era. That is, matter accretion is much more effective than radiation accretion in the mass accretion of PBHs. Such a thing was to be expected and was consistent with our imaginations. Furthermore, we compared the model we obtained for augmentation with the works of Ricotti {\it et al.} and Kamionkowski {\it et al.} in Fig.~(3) to ensure its accuracy.

One of the flaws in the interpretation of the process of evaporation of PBHs was that, in various papers, the apparent horizon of black holes was considered static, and calculations related to Hawking radiation were performed with this assumption. However, it is obvious that PBHs cannot be isolated and there is matter and radiation around them. Especially when we consider them as the constituents of dark matter. We know that the proportionality between the radius and the mass of the Schwarzschild black hole is established, so considering the PBHs as Schwarzschild black holes, it is clear that with the increase in mass, the radius will definitely go out of the static state and become dynamic. Particularly if the increase is continuous, the radius also changes continuously. As a result, there is a competition between mass reduction due to evaporation and mass increase due to accretion.

 Nonetheless, we have to be very careful about evaporation calculations. Hawking radiation is the result of tunneling in the horizon potential barrier. Now, if the horizon is growing, this potential barrier is no longer the same as the static horizon potential barrier. Knowing that the quantum fluctuations on the horizon are related to the Planck length, it is enough to check the graphs related to the rate of change of the radius of black holes over time in order to know in which cases the accretion can cause rapid growth of the radius and as a result, for what mass and at what times accretion can prevent particles from escaping from the black hole's gravity. In this paper, we considered four masses $10^{16}–10^{17} g$, $10^{20}–10^{24} g$, and $1–10^4M_{\odot}$ . The red hatching in these figures means that the growth apparent horizon is so great that it actually forces escaping particles to fall back into PBHs, thus stopping the evaporation of them.
 
This paper is a very interesting start to investigating the models that claim that the evaporation of PBHs creates cosmological effects or that they want to explain a phenomenon with the help of the evaporation of PBHs. It seems that before any calculation to explain a phenomenon with the help of Hawking radiation, it should be checked whether the PBH with a specific mass could have Hawking radiation at all or not.

\section{Acknowledgments}
We are grateful to Pouriya Khaliliyan  for several helpful discussions and comprehensive advice during this work.
\


\begin{thebibliography}{99}
\bibitem{LIGOScientific:2016aoc}
B.~P.~Abbott \textit{et al.} [LIGO Scientific and Virgo],
Phys. Rev. Lett. \textbf{116}, no.6, 061102 (2016)
doi:10.1103/PhysRevLett.116.061102


\bibitem{LIGOScientific:2021djp}
R.~Abbott \textit{et al.} [LIGO Scientific, VIRGO and KAGRA],
[arXiv:2111.03606 [gr-qc]].


\bibitem{Sasaki:2018dmp}
M.~Sasaki, T.~Suyama, T.~Tanaka and S.~Yokoyama,
Class. Quant. Grav. \textbf{35}, no.6, 063001 (2018)
doi:10.1088/1361-6382/aaa7b4


\bibitem{Carr:2020gox}
B.~Carr, K.~Kohri, Y.~Sendouda and J.~Yokoyama,
Rept. Prog. Phys. \textbf{84}, no.11, 116902 (2021)
doi:10.1088/1361-6633/ac1e31



\bibitem{Green:2020jor}
A.~M.~Green and B.~J.~Kavanagh,
J. Phys. G \textbf{48}, no.4, 043001 (2021)
doi:10.1088/1361-6471/abc534



\bibitem{Hutsi:2020sol}
G.~H\"utsi, M.~Raidal, V.~Vaskonen and H.~Veerm\"ae,
JCAP \textbf{03}, 068 (2021)
doi:10.1088/1475-7516/2021/03/068

\bibitem{DeLuca:2021wjr}
V.~De Luca, G.~Franciolini, P.~Pani and A.~Riotto,
JCAP \textbf{05}, 003 (2021)
doi:10.1088/1475-7516/2021/05/003


\bibitem{Franciolini:2021tla}
G.~Franciolini, V.~Baibhav, V.~De Luca, K.~K.~Y.~Ng, K.~W.~K.~Wong, E.~Berti, P.~Pani, A.~Riotto and S.~Vitale,
Phys. Rev. D \textbf{105}, no.8, 083526 (2022)
doi:10.1103/PhysRevD.105.083526


\bibitem{Niemeyer:1999ak}
J.~C.~Niemeyer and K.~Jedamzik,
Phys. Rev. D \textbf{59}, 124013 (1999)
doi:10.1103/PhysRevD.59.124013


\bibitem{Shibata:1999zs}
M.~Shibata and M.~Sasaki,
Phys. Rev. D \textbf{60}, 084002 (1999)
doi:10.1103/PhysRevD.60.084002


\bibitem{Allahyari:2016osl}
A.~Allahyari, J.~T.~Firouzjaee and A.~A.~Abolhasani,
JCAP \textbf{06}, 041 (2017)
doi:10.1088/1475-7516/2017/06/041


\bibitem{Musco:2020jjb}
I.~Musco, V.~De Luca, G.~Franciolini and A.~Riotto,
Phys. Rev. D \textbf{103}, no.6, 063538 (2021)
doi:10.1103/PhysRevD.103.063538


\bibitem{Firouzjaee:2014zfa}
J.~T.~Firouzjaee and G.~F.~R.~Ellis,
Gen. Rel. Grav. \textbf{47}, no.2, 6 (2015)
doi:10.1007/s10714-014-1848-2


\bibitem{Firouzjaee:2015bqa}
J.~T.~Firouzjaee and G.~F.~R.~Ellis,
Phys. Rev. D \textbf{91}, no.10, 103002 (2015)
doi:10.1103/PhysRevD.91.103002


\bibitem{Firouzjaee:2015wps}
J.~T.~Firouzjaee and G.~F.~R.~Ellis,
Eur. Phys. J. C \textbf{76}, no.11, 620 (2016)
doi:10.1140/epjc/s10052-016-4479-9

\bibitem{Hawking:1975vcx}
S.~W.~Hawking,
Commun. Math. Phys. \textbf{43}, 199-220 (1975)
[erratum: Commun. Math. Phys. \textbf{46}, 206 (1976)]
doi:10.1007/BF02345020


\bibitem{Tabasi:2021cxo}
S.~S.~Tabasi and J.~T.~Firouzjaee,
[arXiv:2112.02818 [astro-ph.CO]].


\bibitem{Ricotti:2007au}
M.~Ricotti, J.~P.~Ostriker and K.~J.~Mack,
Astrophys. J. \textbf{680}, 829 (2008)
doi:10.1086/587831
[arXiv:0709.0524 [astro-ph]].


\bibitem{Carr:2009jm}
B.~J.~Carr, K.~Kohri, Y.~Sendouda and J.~Yokoyama,
Phys. Rev. D \textbf{81}, 104019 (2010)
doi:10.1103/PhysRevD.81.104019
[arXiv:0912.5297 [astro-ph.CO]].


\bibitem{Rice:2017avg}
J.~R.~Rice and B.~Zhang,
JHEAp \textbf{13-14}, 22-31 (2017)
doi:10.1016/j.jheap.2017.02.002
[arXiv:1702.08069 [astro-ph.HE]].

\bibitem{Villanueva-Domingo:2021spv}
P.~Villanueva-Domingo, O.~Mena and S.~Palomares-Ruiz,
Front. Astron. Space Sci. \textbf{8}, 87 (2021)
doi:10.3389/fspas.2021.681084
[arXiv:2103.12087 [astro-ph.CO]].


\bibitem{Nayak:2011sk}
B.~Nayak and M.~Jamil,
Phys. Lett. B \textbf{709}, 118-122 (2012)
doi:10.1016/j.physletb.2012.02.010
[arXiv:1107.2025 [gr-qc]].

	\bibitem{hawking-75}
	S W Hawking (1975)
	Comm Math Physics \textbf{43}: 199-220.

	\bibitem{Cheek:2021odj}
A.~Cheek, L.~Heurtier, Y.~F.~Perez-Gonzalez and J.~Turner,
Phys. Rev. D \textbf{105}, no.1, 015022 (2022)
doi:10.1103/PhysRevD.105.015022
[arXiv:2107.00013 [hep-ph]].

\bibitem{Hooper:2020otu}
D.~Hooper and G.~Krnjaic,
Phys. Rev. D \textbf{103}, no.4, 043504 (2021)
doi:10.1103/PhysRevD.103.043504
[arXiv:2010.01134 [hep-ph]].

\bibitem{Keith:2020jww}
C.~Keith, D.~Hooper, N.~Blinov and S.~D.~McDermott,
Phys. Rev. D \textbf{102}, no.10, 103512 (2020)
doi:10.1103/PhysRevD.102.103512
[arXiv:2006.03608 [astro-ph.CO]].

Phys. Rev. D \textbf{81}, 104019 (2010)
doi:10.1103/PhysRevD.81.104019
[arXiv:0912.5297 [astro-ph.CO]].


\bibitem{Chisholm:2011kn}
J.~R.~Chisholm,
Phys. Rev. D \textbf{84}, 124031 (2011)
doi:10.1103/PhysRevD.84.124031
[arXiv:1110.4402 [astro-ph.CO]].


	\bibitem{Bondi:1952ni}
H.~Bondi,
Mon. Not. Roy. Astron. Soc. \textbf{112}, 195 (1952)
doi:10.1093/mnras/112.2.195


\bibitem{Ali-Haimoud:2016mbv}
Y.~Ali-Ha\"\i{}moud and M.~Kamionkowski,
Phys. Rev. D \textbf{95}, no.4, 043534 (2017)
doi:10.1103/PhysRevD.95.043534
[arXiv:1612.05644 [astro-ph.CO]].
\bibitem{Brandenberger:2012zb}
R.~H.~Brandenberger,
[arXiv:1206.4196 [astro-ph.CO]].

\bibitem{Allahverdi:2005fq}
R.~Allahverdi and A.~Mazumdar,
[arXiv:hep-ph/0505050 [hep-ph]].


\bibitem{Rubakov:2017xzr}
V.~A.~Rubakov and D.~S.~Gorbunov,
World Scientific, 2017,
ISBN 978-981-320-987-9, 978-981-320-988-6, 978-981-322-005-8
doi:10.1142/10447


\bibitem{Yang:2021agk}
Y.~Yang,
Mon. Not. Roy. Astron. Soc. \textbf{508}, no.4, 5709-5715 (2021)
doi:10.1093/mnras/stab2966
[arXiv:2110.06447 [astro-ph.CO]].


\bibitem{Mena:2019nhm}
O.~Mena, S.~Palomares-Ruiz, P.~Villanueva-Domingo and S.~J.~Witte,
Phys. Rev. D \textbf{100}, no.4, 043540 (2019)
doi:10.1103/PhysRevD.100.043540
[arXiv:1906.07735 [astro-ph.CO]].

\bibitem{DeLuca:2020fpg}
V.~De Luca, G.~Franciolini, P.~Pani and A.~Riotto,
Phys. Rev. D \textbf{102}, no.4, 043505 (2020)
doi:10.1103/PhysRevD.102.043505
[arXiv:2003.12589 [astro-ph.CO]].



\bibitem{Page:1976wx}
D.~N.~Page and S.~W.~Hawking,
Astrophys. J. \textbf{206}, 1-7 (1976)
doi:10.1086/154350


\bibitem{Niikura:2019kqi}
H.~Niikura, M.~Takada, S.~Yokoyama, T.~Sumi and S.~Masaki,
Phys. Rev. D \textbf{99}, no.8, 083503 (2019)
doi:10.1103/PhysRevD.99.083503
[arXiv:1901.07120 [astro-ph.CO]].


\bibitem{Inoue:2017csr}
Y.~Inoue and A.~Kusenko,
JCAP \textbf{10}, 034 (2017)
doi:10.1088/1475-7516/2017/10/034
[arXiv:1705.00791 [astro-ph.CO]].


\bibitem{Raidal:2017mfl}
M.~Raidal, V.~Vaskonen and H.~Veerm\"ae,
JCAP \textbf{09}, 037 (2017)
doi:10.1088/1475-7516/2017/09/037
[arXiv:1707.01480 [astro-ph.CO]].


\bibitem{Kohri:2014lza}
K.~Kohri, T.~Nakama and T.~Suyama,
Phys. Rev. D \textbf{90}, no.8, 083514 (2014)
doi:10.1103/PhysRevD.90.083514
[arXiv:1405.5999 [astro-ph.CO]].


\bibitem{Carr:1997cn}
B.~J.~Carr and M.~Sakellariadou,
Astrophys. J. \textbf{516}, 195-220 (1999)
doi:10.1086/307071


\bibitem{Carr:2021bzv}
B.~Carr and F.~Kuhnel,
SciPost Phys. Lect. Notes \textbf{48}, 1 (2022)
doi:10.21468/SciPostPhysLectNotes.48
[arXiv:2110.02821 [astro-ph.CO]].

\bibitem{Jacobson:1993hn}
T.~Jacobson,
Phys. Rev. D \textbf{48}, 728-741 (1993)
doi:10.1103/PhysRevD.48.728
[arXiv:hep-th/9303103 [hep-th]].


\end{thebibliography}
\end{document}